# High-sensitivity nanoparticle detection based on the real-splitting indirectly coupled Anti-Parity time symmetric WGMs

Wenxiu Li[1], Hao Zhang[2], Peng Han[1], Xiaoyang Chang[1], Shuo Jiang[1], Yang Zhou[2], Anping Huang[1] and Zhisong Xiao[*1]

[1] *Key Laboratory of Micro-nano Measurement, Manipulation and Physics (Ministry of Education), School of Physics, Beihang University, Beijing 100191, China*

[2] *Research Institute of Frontier Science, Beihang University, Beijing, 100191, China*

zsxiao@buaa.edu.cn

Detecting the size of single nanoparticle with high precision is crucial to understanding the characteristic of the nanoparticle. In this paper, we research the single particle detection based on the Anti-parity time (APT) symmetric indirectly coupled WGMs. The results show that the Anti-parity time symmetric WGM nanoparticle sensor exhibits giant enhancement in frequency splitting compared with single WGM sensor, when the system operating at exceptional point (EP). With respect to the parity-time (PT) symmetric nanoparticle sensor, our research exhibits a real eigenfrequency splitting, which can be directly detected.

## 1. Introduction

In Parity-time symmetry, a non-hermitian Hamiltonian H, defined as [PT, H] =0, can exhibit an entirely real eigenenergy spectra under the PT broken phase, which has attracted much attention in recent years [1-4]. In optical systems, optical microcavities provide an ideal platform for researching non-hermitian physics, because the complex potentials in non-Hermitian quantum systems can be easily realized, through tuning gain or loss rates of the microcavity [3,4]. Most of the novel phenomenon of non-Hermitian systems are found at the exceptional point (EP), in particularly, non-Hermitian systems operating at EP can pave a new way to enhance sensitivity [5-8]. Instead of the traditional linear reaction, in non-Hermitian configurations, the eigenfrequency splitting follows a M root $\Delta\omega \in \varepsilon^{1/M}$ (0<ε<1), representing that the sensitivity of the system will increase with the M orders at the EP, where M is numbers of eigenvalues coalesce at EP and the ε is the perturbation strength. Recently, the second-order and high-order EP are observed in parity–time symmetric optical microcavity systems and Hossein Hodaei has investigated corresponding eigenstates coalesce to achieve higher sensitivity that is proportional to the cube root of the perturbation [9].

Now, APT symmetric configurations, which obey the {PT, H} = 0, also has drawn interests [10-16]. Different from the PT symmetry, the optical systems with APT symmetry should satisfy n(x)=-n$^*$(-x) and dissipative coupling. Anti-Parity time symmetry has been investigated in multi-waveguides [11,12], electrical circuit resonators [13], balanced positive−negative index multilayers [14] and cold-atom lattices [17]. It is worth noting that APT symmetric systems also exist exceptional points [11-13], while there are very limited works on EP in APT symmetric configurations is used to enhance the sensitivity, Martino. De Carlo firstly investigated the real-splitting APT symmetric microscale optical gyroscope [17].

Traditionally, the counterpropagating modes clockwise (CW) mode and counterclockwise (CCW) mode, which have a degenerate frequency, can coexist in WGM microcavity. When a nanoparticle enters the optical mode of the cavity, the interaction between evanescent and nanoparticle will lift the degeneracy, causing frequency splitting [18-23]. The mode splitting introduced by the nanoparticle in principle is proportional to the coupled strength of perturbation ε, which limits the sensitivity and detection limit of WGM nanoparticle sensor. Interestingly, optical microcavity operating at exceptional points can be used for enhancing biosensing [24]. The second-order EP system in parity-time symmetry has square-root at the complex frequency, therefore, eigenfrequency splitting induced by a perturbation ε is proportional to $(i\varepsilon)^{1/2}$ when is utilized for nanoparticle detection [25]. However, the eigenfrequency splitting in parity-time symmetric system induced by the perturbation is complex, which is difficult for the signal read out.

In this work, we used a structure, the indirectly coupled WGM microcavities through two common waveguides, to realize the APT symmetric condition. The APT nanoparticle sensor operating at the EP exhibits a real frequency splitting and a square-root dependence of the sensitivity on the perturbation, and the sensitivity can be giant enhancement compared to single WGM microcavity. With respect to the generally complex eigenfrequency splitting in Parity-time symmetric nanoparticle sensor, APT symmetric sensor can exhibit a real eigenfrequency splitting and this configuration can be achieved more easily constructed than the requirement on balancing the gain and loss in PT symmetric coupled microcavities.

## 2. APT Symmetric single nanoparticle sensor

In the case of PT, eigenvalues of non-hermitian Hamiltonians can be entirely real, i.e. $[PT, H]=0$, while for Anti-parity time symmetric condition, the Hamiltonian should satisfy the condition $\{PT, H\}=0$ [9,18]. Generally, we use an active cavity directly couples a passive cavity with a balanced gain and loss to realize the PT system, and both the cavities have the same resonance frequency. Here, we use two indirectly coupled WGMs through the common waveguides to realize the APT showed in Fig.1, which satisfy the APT condition $\Delta_1=-\Delta_2$, $\gamma_1=\gamma_2$ and $\kappa_1=-\kappa_2^*$.

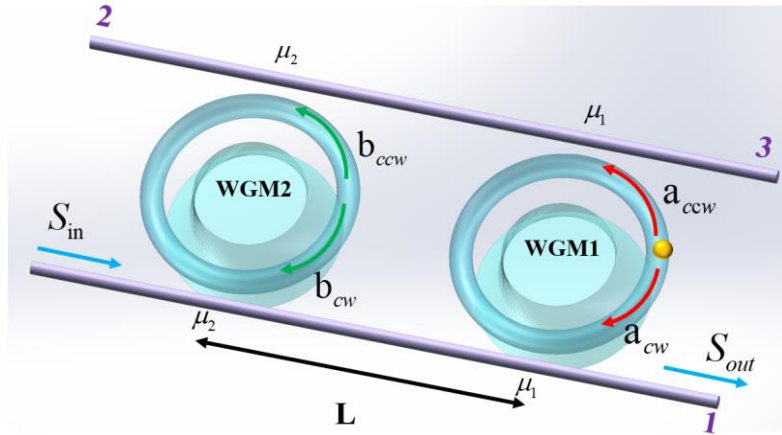

**Fig.1 The schematic of the Anti-parity time symmetric indirectly coupled microresonators sensor, the WGM1 and the WGM2 coupled through two common waveguides. Mode $a_{cw}$ and $b_{cw}$ are induced by a nanoparticle.**

When a nanoparticle enters the mode of the WGM, the optical evanescent fields in the WGM will be scattered by particle. The mode $a_{cw}$ and $a_{ccw}$ are coupled by backscattering in WGM1 and they indirectly coupled with mode $b_{cw}$ and $b_{ccw}$ in WGM2 through the common waveguide, respectively. When we set the system at EP, a nanoparticle with perturbation strength g is introduced into the optical mode. The perturbation non-Hermitian Hamiltonian in the traveling wave of the single

nanoparticle can be written as

$$H = H_0 + H_1 + H_2$$
$$H_0 = \hbar\omega_a a_{cw}^\dagger a_{cw} + \hbar\omega_a a_{ccw}^\dagger a_{ccw} + \hbar\omega_b b_{cw}^\dagger b_{cw} + \hbar\omega_b b_{ccw}^\dagger b_{ccw}$$
$$H_1 = \sum_{m,m'=cw,ccw} \hbar g a_m^\dagger a_{m'} \quad (1)$$
$$H_2 = \sum_{m,m'=cw,ccw} -i\hbar(\gamma_s + \gamma_a) a_m^\dagger a_{m'}$$
$$H_3 = \hbar(\kappa_1 a_{cw}^\dagger b_{cw} + \kappa_2 b_{cw}^\dagger a_{cw}) + \hbar(\kappa_1 a_{ccw}^\dagger b_{ccw} + \kappa_2 b_{ccw}^\dagger a_{ccw})$$

where, $H_0$ is the free Hamiltonian for the APT symmetric indirectly coupled system, $H_1$ corresponds to the interaction between the clockwise(cw) and counterclockwise(ccw) modes induced by the nanoparticle. $H_2$ is the cavity loss due to the particle scattering and absorption with the coefficient $\gamma_s$ and $\gamma_a$, respectively. In practice, the backscattering strength is $g = -\text{Re}[\alpha] f^2(\vec{r}) \omega_c / 2V_m$, where $f(r)$ is the cavity mode fuction, $V_m$ is the mode volume of cavity and $\alpha = 4\pi R^3 \varepsilon_1 (\varepsilon_p - \varepsilon_1)/(\varepsilon_p + 2\varepsilon_1)$ is the polarizability of a particle with $\varepsilon_p$ and $\varepsilon_l$ denote dielectric permittivities of the nanoparticle and the environment medium, respectively. $\gamma_s$ and $\gamma_a$ can be neglected when the size of the particle is small [19]. $H_3$ is the indirectly modes coupling between the two WGM cavities. Thereby, the corresponding eigenfrequencies in APT system are

$$\omega_{APT,1} = \frac{\omega_a + \omega_b}{2} - i\gamma + \sqrt{\Delta^2 + \kappa_c^2},$$
$$\omega_{APT,2} = \frac{\omega_a + \omega_b}{2} - i\gamma - \sqrt{\Delta^2 + \kappa_c^2},$$
$$\omega_{APT,3} = \frac{\omega_a + \omega_b}{2} - i\gamma + g + \sqrt{\Delta^2 + g^2 + 2g\Delta + \kappa_c^2}, \quad (2)$$
$$\omega_{APT,4} = \frac{\omega_a + \omega_b}{2} - i\gamma + g - \sqrt{\Delta^2 + g^2 + 2g\Delta + \kappa_c^2},$$

where, $\Delta = (\omega_a - \omega_b)/2$ is the effective detuning between the two WGMs. Eq. (2) shows that the nanoparticle lifts the eigenfrequency degeneracy of the supermodes in which two supermodes experience a frequency shift and linewidth change, while the other two supermodes are not affected by the nanoparticle serving as reference signals. The indirectly coupled strength is $\kappa_{1,2} = \mu_1 \mu_2 e^{-i\theta_{1,2}}$, and the $\mu_{1,2}$ is the real coupling between the waveguide and the WGM cavities and the $\theta_{1,2} = 2\pi n_g L/\lambda$. The system will be APT criteria when $\kappa_1 = -\kappa_2^* = \kappa_c$, where a possible solution is given by $\theta_{1,2} = (2m+1)\frac{\pi}{2}$, ($m \in \mathbb{N}$). In experiment, we can choose appropriate L to implement that solution, therefore, $\kappa_c = i\mu_1\mu_2$.

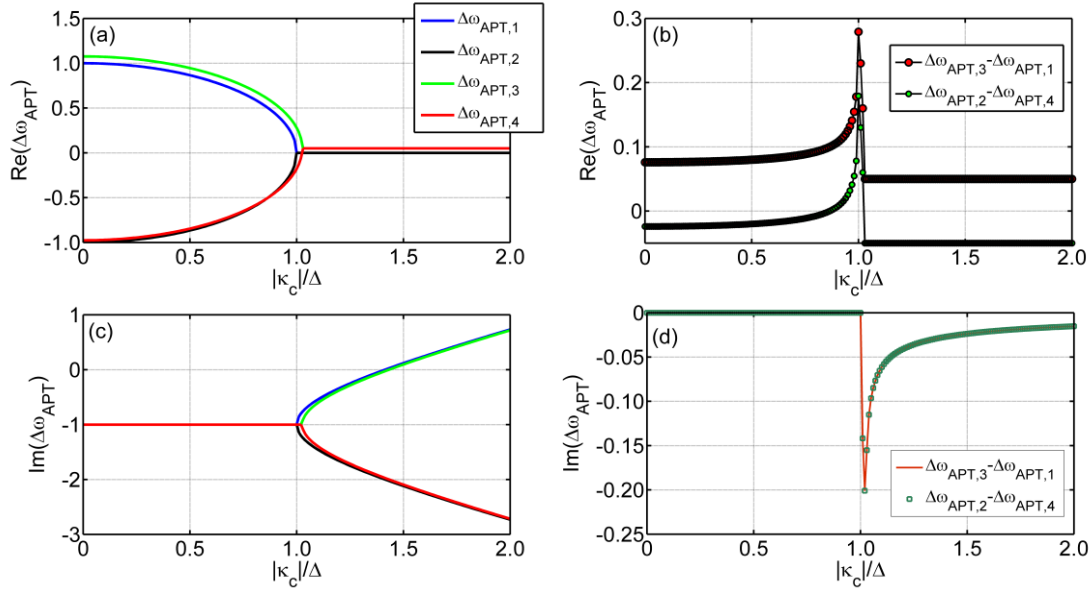

**Fig.2 (a) and (c) The normalized real part and the image part of the eigenfrequencies in APT indirectly coupled resonators varies with the coupled coefficient κ when a single nanoparticle enters to the WGM1, respectively. (b) and (d) The values of the normalized real and image frequency splitting between the perturbed and reference supermodes with the changing of coupling strength κ.**

Fig.2 (a) and Fig.2(c) depict that the eigenmodes take the degenerate resonance frequency but different decay rates in the APT symmetric phase ($|\kappa_c| > \Delta$) and the eigenmodes take nondegenerate resonance frequency but the same decay rate in the APT symmetric broken phase ($|\kappa_c| < \Delta$). The maximal real splitting $\Delta\omega_{APT}$ between perturbed supermodes ($\omega_{APT,3}$ and $\omega_{APT,4}$) and reference supermodes ($\omega_{APT,1}$ and $\omega_{APT,2}$) appear on the EP due to the square-root topology and in this point the imaginary part splitting is zero for small particle are displayed in Fig.2(b) and Fig.2(d), respectively. With the increasing of the coupled coefficient, the APT system will come from the APT symmetric broken phase to the APT symmetric phase, the splitting will not increase. The $|\omega_a - \omega_b| = |2\kappa_c|$ is the exceptional point of the APT indirectly coupled resonators through adjusting the coupling strength between the WGM and the waveguide, we can make the system locate at the EP. In addition, we need make the same effective loss rate in each resonator, i.e., $\gamma=\gamma_{1,ext}-2\mu_1^2-\alpha_1=\gamma_{2,ext}-2\mu_2^2-\alpha_2$ where $\gamma_{1,ext}$ ($\gamma_{2,ext}$) are the external gains and $\alpha_1$ ($\alpha_2$) represent the intrinsic losses in resonator and $\gamma$ is needed to be negative [17]. Finally, the Heisenberg dynamics equation describing the optical modes in the APT indirectly coupled system is

$$\dot{a}_{cw}=-i(\omega_a+g)a_{cw}-\gamma a_{cw}-iga_{ccw}-ik_c b_{cw},$$
$$\dot{a}_{ccw}=-i(\omega_a+g)a_{ccw}-\gamma a_{ccw}-iga_{cw}-ik_c b_{ccw}-i\mu_1 e^{i\theta}s_{in},$$
$$\dot{b}_{cw}=(-i\omega_b-\gamma)b_{cw}-ik_c a_{cw},$$
$$\dot{b}_{ccw}=(-i\omega_b-\gamma)b_{ccw}-ik_c a_{ccw}-i\mu_2 s_{in},$$
(3)

The transmission spectrum $s_{out}/s_{in}$ in port3 showed in Fig.1 can be obtained through the relation $s_{out}=i\mu_1 a_{ccw}+i\mu_2 e^{i\theta}b_{ccw}$.

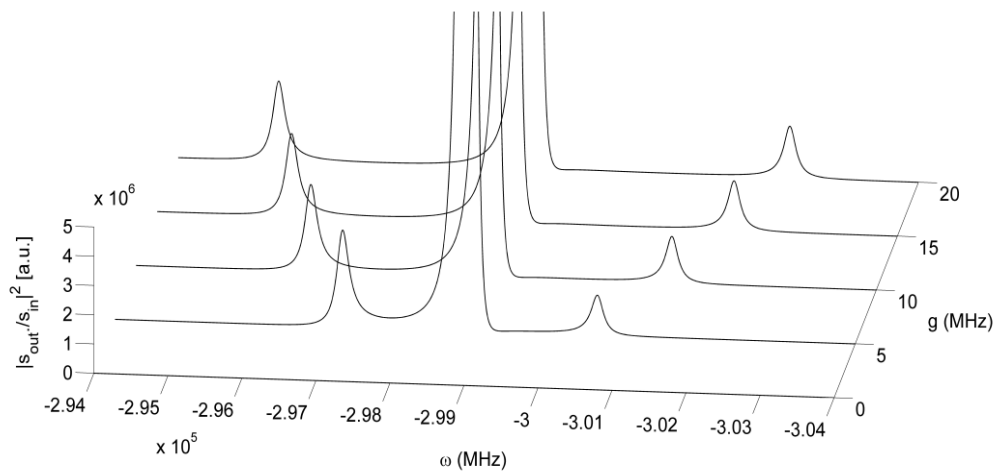

Fig.3 The output spectrum operating at the exceptional point of an APT symmetric nanoparticle sensor with the changing of the perturbation strength g. Here, $R_1$= 20 μm, $R_2$ =19.95 μm, γ=-100 MHz, λ =1.55 μm.

From Fig.3, the nanoparticle inducing eigenfrequency splitting will move from the exceptional point to the broken phase. The middle peak is the reference signal that remains original resonance frequency. Intuitively, two supermodes are not affected by the nanoparticle and have two collapsing eigenfrequencies to make the middle peak much higher. The other two eigenfrequencies are affected by the particle, the more they depart from each other, the less they influence each other, thus the peaks decrease.

When the system located at the EP, according to Eq. (3), due to the nanoparticle is very small, i.e., g<<Δ , the frequency splitting can be written as

$$\omega_{APT1,2} = \frac{\omega_a - \omega_b}{2} - i\gamma + g \pm \sqrt{g^2 + 2g\mu_1\mu_2} \qquad (4)$$

which is proportional to the square-root of the coupled coefficient. For a single WGM nanoparticle sensor with the same perturbation g, the eigenfrequency splitting is $\Delta\omega_{SINGLE} = 2g$, therefore, the sensitivity enhancement factor S can be defined as

$$S = \frac{\Delta\omega_{APT}}{\Delta\omega_{SINGLE}} = \sqrt{\frac{\xi^2 v_g}{4g\pi R_{1,2}}} \qquad (5)$$

Based on the Ref. [17], that coupled coefficient between the WGM with the common waveguide is $\mu_{1,2}^2 = \xi_{1,2}^2 v_g / (2\pi R_{1,2})$, $\xi$ is the fraction of the power coupled from the waveguide to the WGM, $R_{1,2}$ is the radius of the cavity and $v_g$ is the group velocity.

For the Parity-time symmetry sensor, the frequency difference $\Delta\omega_{PT} = \sqrt{-i\gamma g}$ is complex splitting [25]. The APT nanoparticle sensor we proposed can exhibit a real splitting, comparing to the complex splitting in the Parity-time symmetric sensor, which means that the eigenfrequency splitting can be directly detected. Fig.4 shows the enhancement of the splitting of an APT sensor at the EP with respect to a single WGM sensor. The sensitivity enhancement of three orders of magnitude at the exceptional point.

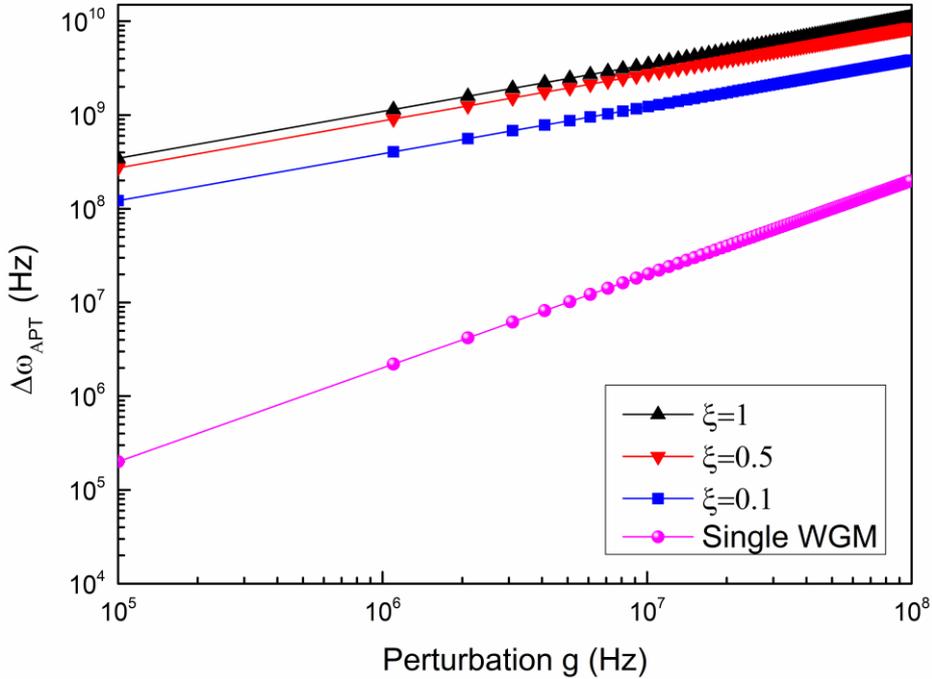

**Fig.4 Comparison the eigenfrequency splitting at the EP between the APT nanoparticle sensor and the single WGM splitting. The parameters are set as $R_1$=20μm, $R_2$=19.95μm, $\xi$=0.1, 0.5, 1.**

## 3. The detection limit of the APT Symmetric nanoparticle sensor

In designing nanoparticle sensors with large sensitivity, one important consideration is the detecting precision. Now, we research the detection limit of the APT symmetric nanoparticle sensor. First, we introduce the frequency splitting quality $Q_{sp}$ [24],

$$Q_{sp} = \frac{\text{Re}(\omega_{APT,+}) - \text{Re}(\omega_{APT,-})}{-\text{Im}(\omega_{APT,+}) - \text{Im}(\omega_{APT,-})} \qquad (6)$$

If the $Q_{sp} > 1$, the eigenfrequency splitting can be resolved easily in experiment, when $Q_{sp} = 1$, the detection limit $g_{min}$ can be obtained. We chose $\omega_{APT,3}$ as $\omega_{APT,+}$ and the $\omega_{APT,1}$ as the $\omega_{APT,+}$, because $\omega_{APT,1}$ corresponding supermodes will not experiences a frequency shift and linewidth change can serve as reference signals. Thus, the detection limit of the perturbation g is

$$g_{min} = \frac{\pi R_{1,2} \omega_c^2}{4Q^2 \xi^2 v_g} \qquad (7)$$

where Q is the quality factor of the WGM. The theoretical detection limit of the perturbation g is inversely proportional to the indirectly coupled coefficient between the two cavities and Q factor. Therefore, in Fig.5, we consider the experiment parameters $V_m \sim 500$ μm$^3$, f (r) =0.26, $\varepsilon_p = 1.592^2$ for polystyrene (PS) nanoparticle and $\varepsilon_l = 1^2$ for air [26]. The theoretical detection limit for PS nanoparticle reaches to 2.95 nm when Q= $5 \times 10^6$, and the higher Q factor is, the better detection limit $R_{min}$ will be obtained.

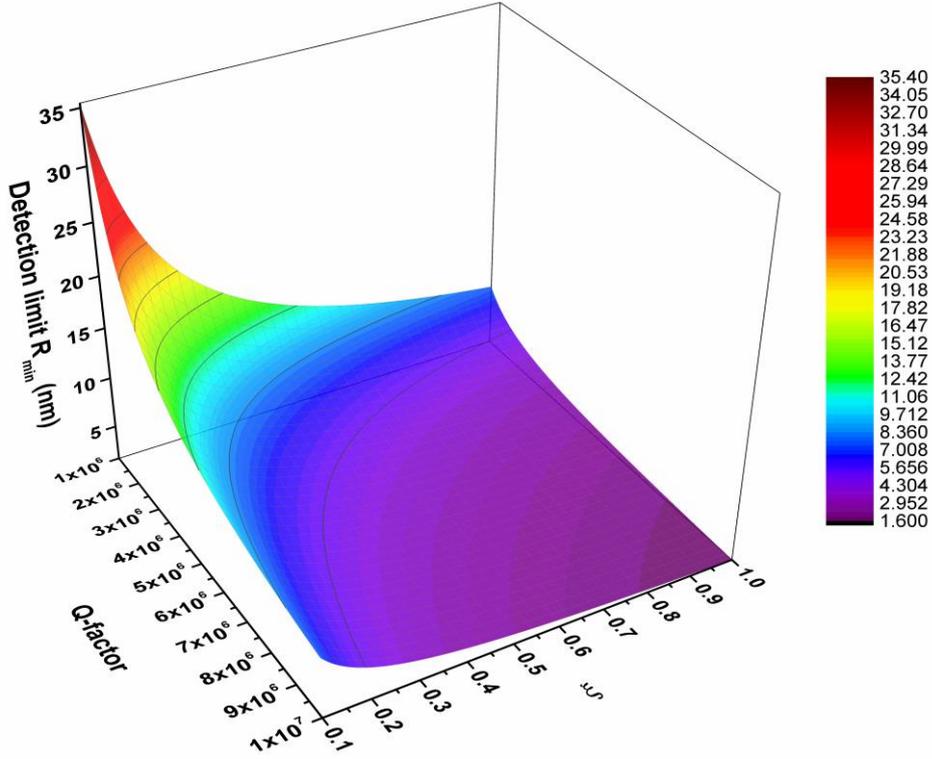

Fig.5 The detection limit of the size $R_{min}$ varies with different coupling strength and Q-factor at exceptional point. Here, $R_1=20\mu m$, $R_2=19.95\mu m$, $\lambda=1550nm$.

## 4. APT Symmetric multiple nanoparticles

For detecting the multiple nanoparticles, we need reconsider the perturbation strength and relative position of each nanoparticle in the Hamiltonian. The scattering strength of light from $a_{cw}$ mode to $a_{ccw}$ mode is determined by the interference of scattered light induced by each nanoparticle. The revisionary interaction Hamiltonian induced by the nanoparticle is

$$H_{1,muti} = \sum_{i=1}^{N} \hbar g_i (a_{cw}^\dagger a_{cw} + a_{ccw}^\dagger a_{ccw} + a_{cw}^\dagger a_{ccw} e^{-2im\beta_i} + a_{ccw}^\dagger a_{cw} e^{2im\beta_i})$$
$$H_{2,muti} = \sum_{i=1}^{N} -i\hbar(\gamma_{a,i} + \gamma_{s,i})(a_{cw}^\dagger a_{cw} + a_{ccw}^\dagger a_{ccw} + a_{cw}^\dagger a_{ccw} e^{-2im\beta_i} + a_{ccw}^\dagger a_{cw} e^{2im\beta_i})$$

(8)

where $\beta_i$ is the angular position of the i-th nanoparticle and m is the azimuthal number of the mode. Assuming the perturbation strength of each particle is same, the corresponding eigenfrequencies are

$$\omega_{muti,1} = \frac{1}{2}\left(g(N-\sqrt{xy})-2i\gamma-\sqrt{-4\kappa_c^2+(g(N-\sqrt{xy})+2\Delta)^2}\right),$$
$$\omega_{muti,2} = \frac{1}{2}\left(g(N-\sqrt{xy})-2i\gamma+\sqrt{-4\kappa_c^2+(g(N-\sqrt{xy})+2\Delta)^2}\right),$$
$$\omega_{muti,3} = \frac{1}{2}\left(g(N-\sqrt{xy})-2i\gamma-\sqrt{-4\kappa_c^2+(g(N+\sqrt{xy})+2\Delta)^2}\right),$$
$$\omega_{muti,4} = \frac{1}{2}\left(g(N-\sqrt{xy})-2i\gamma+\sqrt{-4\kappa_c^2+(g(N+\sqrt{xy})+2\Delta)^2}\right).$$
(9)

here, $\Delta = (\omega_a - \omega_b)/2$, $x = \sum_{i=1}^{N} e^{-2im\beta_i}$ and $y = \sum_{i=1}^{N} e^{2im\beta_i}$. The light scattered by the particle forming a symmetric mode (SM) and an asymmetric mode (ASM), since the nanoparticle locates at the node of the ASM mode field, the mode remains original resonance frequency, and mode shift appears when the particle locates at the antinode of the SM mode field. When two or more particles are deposited in the WGM cavity, the reference mode will be affected by the particle, because each of the particle can not ensured to locate at the node of the SM in WGM cavity, thereby, the the mode will not remain original frequency corresponding to the Eq. (9). Fig.6 shows two cases about the two identical nanoparticles with the same perturbation strength （Fig.6(a)）and two different nanoparticles with different perturbation strengths （Fig.6(b)）. The frequency splitting cyclically changes from maximum to minimum, because constructive interference or destructive interference between scattered modes by the two nanoparticles with the varying of angular position of the second particle $\beta_2$. When the perturbation strength is different the minimum of the destructive will not reaches to zero from Fig.6 (b).

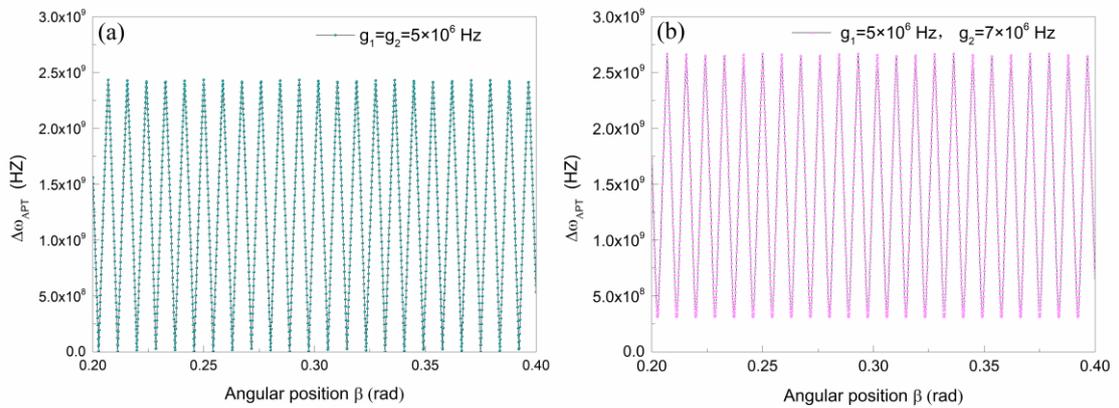

**Fig.6 (a) The two particles are identical with the same perturbation strength g in an APT nanoparticle sensor. (b) The two particles are different with the different perturbation**

**strength g in an APT nanoparticle sensor.**

The the variation of eigenfrequency splitting when N different nanoparticles are deposited one by one at random angular positions on an APT symmetric sensor displayed in Fig.7. The results show that APT symmetric sensor operation in the EP has better sensitivity than a single WGM senor. In the multiparticle condition, the the value of splitting is discrete and not increase linearly with the increasing of the particle numbers due to that two reasons: (i) the frequency splitting impacted by the number and position of the particle simultaneously; and (ii) when the first particle entering the cavity, the APT system will move away from the EP, hence, the second and the after particles will start with a APT system that is not located at EP.

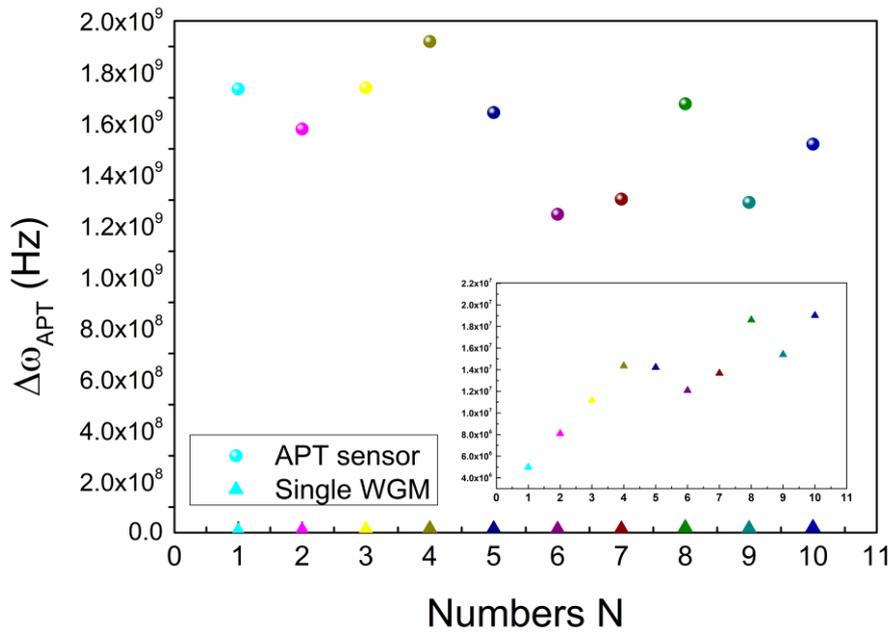

**Fig.7 Numerical results of eigenfrequencies splitting varies with N nanoparticles deposited on the APT sensor (color sphere) and a single WGM (color triangle). The inset is the results of the single WGM sensor. Here, β=（0,0.2,0.62,0.45,0.8,0.73,1.2,1.5,0.9,1.87）×π, g=5MHz.**

## 5. Conclusions

We investigated an integrated Anti-parity symmetric indirectly coupled WGM nanoparticle sensor. As s result, the configuration operating at the exceptional point shows a significant enhancement with respect to the classical nanoparticle sensor, achieving an eigenfrequency splitting the 3 orders of magnitude higher than a single

WGM sensor. Take polystyrene (PS) nanoparticle as an example, the theoretical detection limit reaches to 2.95nm when $Q=5\times10^6$. With regards to the Parity-time symmetric nanoparticle sensor, the Anti-PT-symmetric solution is more suitable for nanoparticle detecting, because it exhibits a real frequency splitting. Compared with balancing the gain and loss in parity-time symmetric nanoparticle senor, APT symmetric sensor just requires to adjusting the detuning between the two WGMs, keeping the configuration at the EP more reliable. We believe that the APT symmetric sensor will pave a way to realize the ultra-high sensitivity integrated optical nanoparticle sensor.

**Acknowledgement.** Financial support from the National Natural Science Foundation of China (11574017, 11574021,11804017,51372008) and Beijing Academy of Quantum Information Sciences(Y18G28).